\documentclass{jetpl}
\twocolumn

\usepackage{amsmath,amssymb,epsfig,color,pstricks,bm,graphics,alltt}

\usepackage[colorlinks,plainpages=false,linkcolor=black,urlcolor=blue,citecolor=black,pdfpagemode=UseNone,pdfstartview=FitBH]{hyperref}

\definecolor{MyDarkBlue}{rgb}{0,  0.3,  0.9}
\definecolor{MyDarkBlack}{rgb}{0,  0,  0}

\begin{document}

\lat

\title{Electronic Structure of New Multiple Band Pt-Pnictide Superconductors APt$_3$P}

\rtitle{Electronic Structure of New Multiple Band Pt-Pnictide Superconductors}

\sodtitle{Electronic Structure of New Multiple Band Pt-Pnictide Superconductors APt$_3$P}

\author{$^a$I.\ A.\ Nekrasov\thanks{E-mail: nekrasov@iep.uran.ru},
$^{a,b}$M.\ V.\ Sadovskii\thanks{E-mail: sadovski@iep.uran.ru}}

\rauthor{I.\ A.\ Nekrasov, M.\ V.\ Sadovskii}

\sodauthor{Nekrasov, Sadovskii }

\sodauthor{Nekrasov, Sadovskii}

\address{$^a$Institute for Electrophysics, Russian Academy of Sciences, 
Ural Branch, Amundsen str. 106,  Ekaterinburg, 620016, Russia\\
$^b$Institute for Metal Physics, Russian Academy of Sciences, Ural Branch,
S.Kovalevskoi str. 18, Ekaterinburg, 620990, Russia}

\dates{May 2012}{*}

\abstract{
We report LDA calculated band structure, densities of states and
Fermi surfaces for recently discovered Pt-pnictide superconductors 
APt$_3$P (A=Ca,Sr,La), confirming their multiple band nature.
Electronic structure is essentially three dimensional, in contrast to 
Fe pnictides and chalcogenides. LDA calculated Sommerfeld coefficient
agrees rather well with experimental data, leaving little space for very strong 
coupling superconductivity, suggested by experimental data on specific heat
of SrPt$_3$P. Elementary estimates show, that the values of critical temperature
can be explained by rather weak or moderately strong coupling, while the
decrease of superconducting transition temperature $T_c$ from Sr to La compound
can be explained by corresponding decrease of total density of states at the 
Fermi level $N(E_F)$. The shape of the density of states near the Fermi level 
suggests that in SrPt$_3$P electron doping (such as replacement Sr by La) 
decreases $N(E_F)$ and $T_c$, while hole doping (e.g. partial 
replacement of Sr with K, Rb or Cs, if possible) would increase $N(E_F)$ and 
possibly $T_c$.
}

\PACS{71.20.-b, 74.70.-b, 74.20.-z, 74.20.Fg}

\maketitle

\section{Introduction}

After the discovery of first iron-pnictide high-temperature superconductors 
\cite{kamihara_08} (see reviews in Refs.~\cite{UFN_90,Hoso_09}) several other 
iron-pnictide and iron-chalcogenide families of superconductors were intensively
studied in recent years (see reviews in Refs.~\cite{UFN_90,Hoso_09} and
pnictide-chalcogenide comparison in Ref. \cite{PvsC}). The search for new
systems produced several new superconductors, which are chemical analogues of 
iron-pnictides and chalcogenides, such as  BaNi$_2$As$_2$ \cite{Bauer08}, 
SrNi$_2$As$_2$ \cite{Ronning08}, SrPt$_2$As$_2$ \cite{Kudo10}, SrPtAs 
\cite{Elgazzar}, but with pretty low superconducting $T_c$.
Recently, another family of Pt-based superconductors with chemical composition
APt$_3$P (A=Sr,Ca,La) \cite{Takayama} was discovered, with experimental values 
of $T_c$ equal to 8.4K, 6.6K and 1.5K correspondingly.
Based on the observation of nonlinear temperature behavior of Hall resistivity, 
the authors of Ref. \cite{Takayama} supposed the multiple band nature of 
superconductivity in these new systems, while specific heat data on SrPt$_3$P
has lead them to a conclusion on very strong coupling nature of 
superconductivity \cite{Takayama}.

Below we present the results on LDA calculated electronic structure for 
SrPt$_3$P and LaPt$_3$P, as well as some comparison with previously studied
Pt-based superconductor SrPt$_2$As$_2$ \cite{SrPtAs}. We also present some
elementary estimates concerning superconductivity in APt$_3$P system.

\section{Crystal structure}

Crystals of SrPt$_3$P system have tetragonal space group {\it P$\frac{4}{n}$mm}
with a=5.8094\AA~and c=5.3833\AA~\cite{Takayama}.
Between Sr layers there are anti perovskite Pt$_6$P octahedra where basal Pt1 
atoms occupy 4e (1/4,1/4,1/2) positions and apical Pt2 occupy 2c (0,1/2,0.1409). 
Phosphorus inside octahedra also occupy 2c position, but with $z$=0.7226.
These octahedra are not ideal -- distance from basal plane to apical Pt2 
atoms is different, while basal Pt1 atoms form perfect squares.
From Fig. 1 it is clear that because of alternating edge sharing Pt$_6$P 
octahedra, basal Pt1 atoms form two dimensional square lattice. 
For LaPt$_3$P we assume the same crystal structure as for SrPt$_3$P.

\begin{figure}[h]
\includegraphics[clip=true,width=0.5\textwidth]{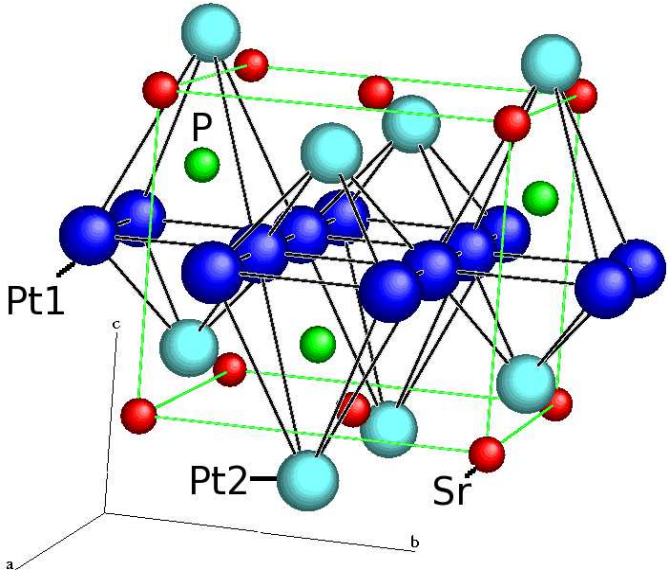}
\caption{Fig. 1. Crystal structure of SrPt$_3$P. Pt$_6$P octahedra are shown
by Pt-Pt bonds.}
\end{figure}

\section{Electronic structure}

Using the experimentally established crystal structure \cite{Takayama} we
performed LDA band structure calculations within the linearized muffin-tin 
orbitals method (LMTO)~\cite{LMTO} with default settings.

In Fig.~2 we present LDA calculated densities of states (DOS) of 
SrPt$_3$P (upper panel) and LaPt$_3$P (lower panel).
Most of spectral weight from -8 eV up to 2 eV belongs to Pt1 (line with circles) 
and Pt2 5d states (line with $\times$). Phosphorus 3p states give rather 
small contribution to the DOS (dot-dashed line).
DOS at the Fermi level is formed mainly by Pt1-5d states
(Pt1 atoms make 2d-square lattice see Fig. 1) with a bit of admixture
of Pt2-5d and P-3p states.

The values of the DOS at the Fermi level $N(E_F)$ for Sr and La compounds
are correspondingly 4.69 states/eV/cell and 3.77 states/eV/cell. These values 
are comparable with those for pnictides with moderate $T_c$ values \cite{ahight}.
Calculating Sommerfeld coefficient $\gamma_b=\frac{\pi^2}{3}N(E_F)$
we obtain 11 mJ/mol/K$^2$ and 8.9 mJ/mol/K$^2$ for SrPt$_3$P and 
LaPt$_3$P respectively. For Sr compound the experimental value of Sommerfeld 
coefficient is $\gamma^{exp}=$12.7 mJ/mol/K \cite{Takayama} and agrees rather 
well with our band structure estimates. In fact $\gamma^{exp}$ should be 
larger than calculated (free electron) $\gamma_b$, because of DOS renormalization 
due to electron-phonon interaction (or interaction with other collective modes):
$\gamma=(1+\lambda)\gamma_b$, where $\lambda$ is corresponding dimensionless
coupling constant. Comparing experimental data and calculated results we
get an estimate of $\lambda$ of the order of 0.15 only, which corresponds to
rather weak coupling and is too small to obtain the experimental values of 
$T_c$ (see below). Note also that experimental estimates \cite{Takayama} of
Wilson ratio produced the values of $R_W\sim 1$, which signifies the absence of
strong correlations in SrPt$_3$P. 

Our results for DOS show, that in SrPt$_3$P the Fermi level is located
just in the middle of the slope of a peak in the DOS, so that electron 
doping decreases $N(E_F)$, while hole doping the other way around increases it.
This may be important for superconductivity (see discussion below).

\begin{figure}[h]
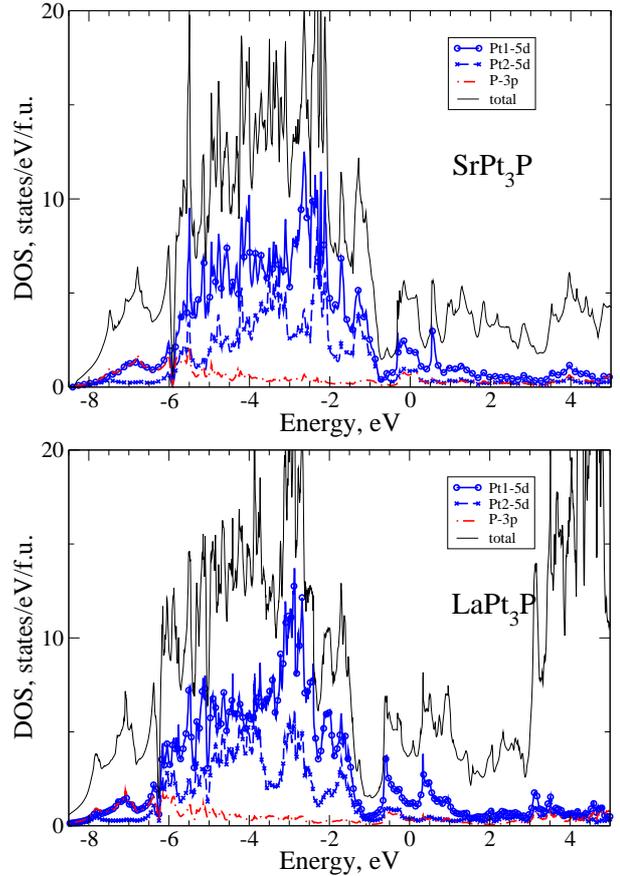

\includegraphics[clip=true,width=0.45\textwidth]{SrPt3P_dos.eps}
\includegraphics[clip=true,width=0.45\textwidth]{LaPt3P_dos.eps}
\caption{Fig. 2. LDA calculated densities of states (DOS) of 
SrPt$_3$P (upper panel) and LaPt$_3$P (lower panel).
Solid black line - total DOS, line with circles - Pt1-5d DOS,
line with $\times$ - Pt2-5d DOS and dot-dashed line - P-3p DOS.
The Fermi level $E_F$ is at zero energy.} 
\end{figure}

In Fig.~2 we show band dispersions of SrPt$_3$P (black lines) and LaPt$_3$P 
(gray lines) in the vicinity of the Fermi level.
These are quite different from band dispersions reported for 
chemically similar material SrPt$_2$As$_2$ \cite{SrPtAs}
and also from iron pnictides and chalcogenides \cite{PvsC}.
First of all, both APt$_3$P compounds are essentially three
dimensional as one can see from dispersions in $\Gamma$-$Z$ direction.
From chemical composition point of view (neglecting lattice relaxation effects) 
LaPt$_3$P is ``electron doped'' SrPt$_3$P system (one extra electron in La).
This results in almost rigid shift of the La compound bands
down in energy for about 0.3 eV with respect to the bands of Sr compound
(see Fig. 2). Close to the Fermi level for LaPt$_3$P there are several band crossings and Van-Hove
singularities. Thus Fermi surface topology can be changed rather easy upon 
doping.

\begin{figure*}
\includegraphics[clip=true,width=0.9\textwidth]{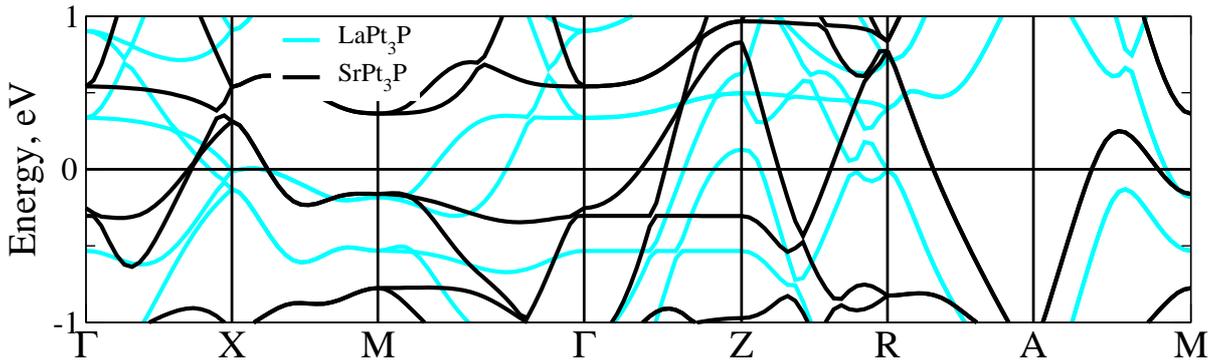}
\caption{Fig. 3. LDA calculated band dispersions in the vicinity of the Fermi 
level for SrPt$_3$P (black line) and LaPt$_3$P (gray line).
The Fermi level is at zero energy.
} 
\end{figure*}

In Fig.~4 and Fig.~5 we present LDA calculated Fermi surfaces (FS) of SrPt$_3$P 
and LaPt$_3$P correspondingly and also separately drawn different FS sheets.
Overall shape of the APt$_3$P FS is very different from those in typical iron 
pnictides or chalcogenides \cite{Nekr2,KFeSe}. First of all it looks pretty 
three-dimensional and does not have any well developed cylinders. 

\begin{figure*}[h]
\includegraphics[clip=true,width=0.7\textwidth]{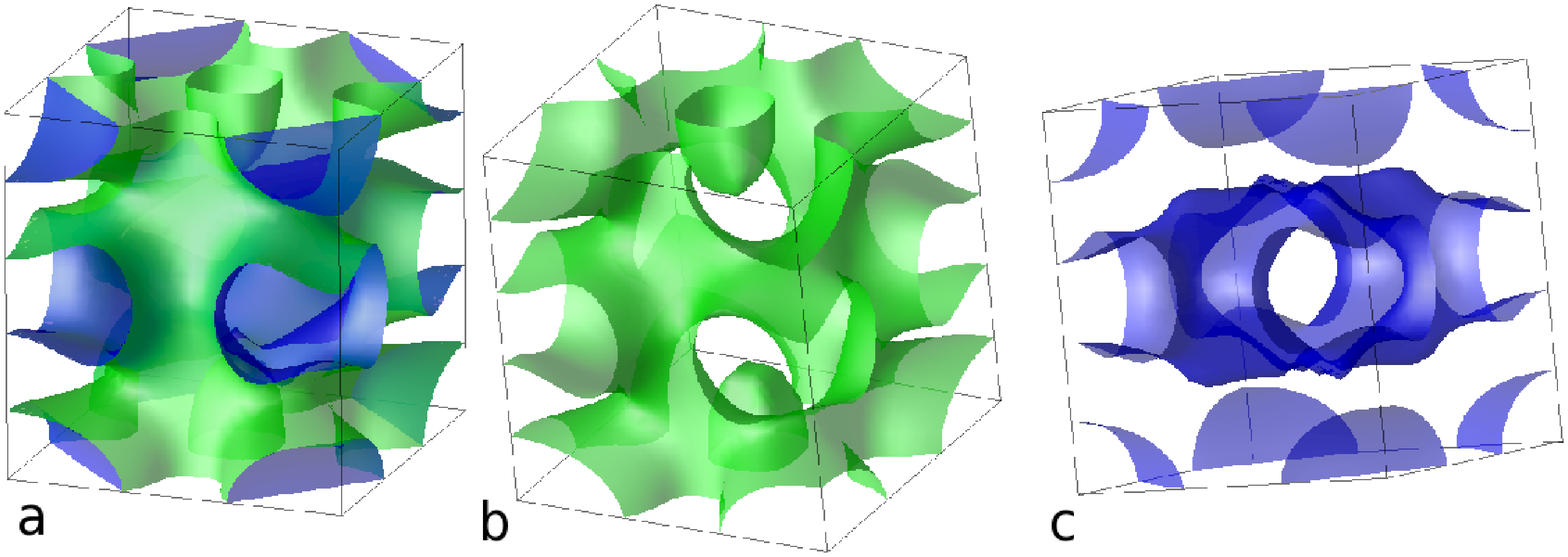}
\caption{Fig. 4. LDA calculated Fermi surface 
for SrPt$_3$P (a) and its separate sheets (b,c).
} 
\end{figure*}

The overall conclusion on electronic structure of APt$_3$P family is that it 
really the new class of multiple band superconductors, as was expected from
Hall measurements in Ref. \cite{Takayama}. SrPt$_3$P system is essentially 
two-band superconductor, while LaPt$_3$ band structure in the vicinity of the
Fermi level is even more complicated. Fermi surfaces of both systems are 
characterized by multiple sheets and pockets in the Brillouin zone, which
produce very complicated topology and variability under doping.

\section{Remarks on Superconductivity}

From the general symmetry analysis \cite{VG,SU} it is known that
in case of tetragonal symmetry and in spin-singlet case we can in principle
observe either the usual isotropic or anisotropic $s$-wave Cooper pairing or 
several types of $d$-wave pairing. Most probable is, of course, the case of
s-wave superconductivity, as proposed in Ref. \cite{Takayama}.
Additional complications arise from multiple band nature of the APt$_3$P
compounds. The three-dimensional multiple sheet FS topology may lead to 
complicated superconducting gap structure, with different energy gaps on 
different FS sheets. The $T_c$ value and gap ratios in the multiple  band 
systems are actually determined by rather complicated interplay of
intraband and interband couplings in Cooper channel, as well as by
partial DOS ratios at different FS sheets \cite{Gork,KS09}.

\begin{figure*}
\includegraphics[clip=true,width=0.9\textwidth]{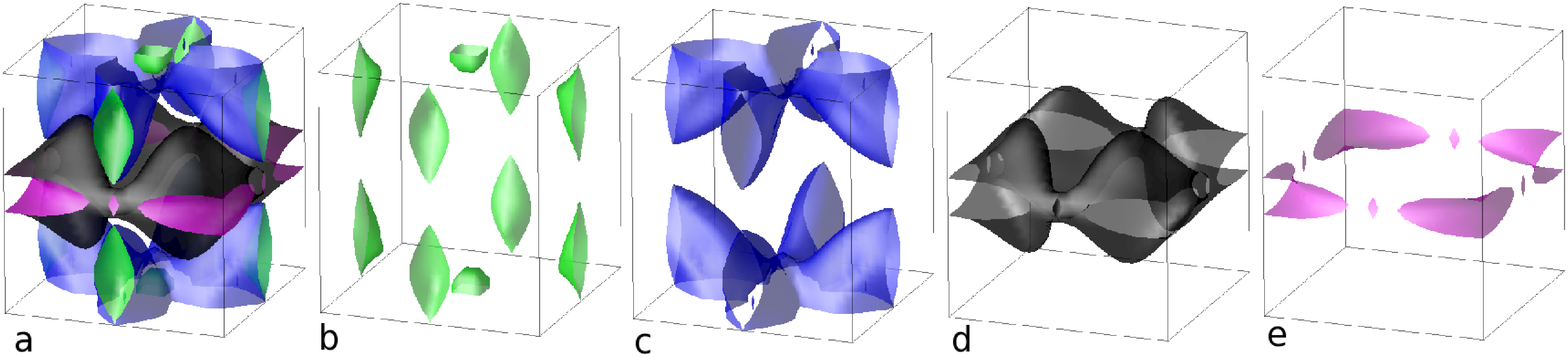}
\caption{Fig. 5. LDA calculated Fermi surface 
for LaPt$_3$P (a) and its separate sheets (b -- e).
} 
\end{figure*}

Nevertheless simple BCS approach allows us to make simple estimates  of
dimensionless coupling constant $\lambda$ value from experimental values of
$T_c$. Using the simple BCS expression $T_c=1.14\omega_De^{-1/\lambda}$ with
experimental value of Debye frequency $\omega_D$=190K \cite{Takayama}
we obtain weak coupling value of $\lambda$=0.31 for $T_c$=8.4K. Reducing this 
value of $\lambda$ proportionally to the decrease of DOS at the Fermi level 
(from 4.69 states/eV/cell in Sr compound to 3.77 states/eV/cell in
La system) we obtain $T_c$=4K for LaPt$_3$P in a reasonable agreement with
experimental value of 1.5K.  

As simple BCS expression for $T_c$ is probably too crude, we also used the
McMillan expression \cite{MM}:
\begin{equation}
T_c=\frac{\omega_D}{1.45}\exp\left(-\frac{1.04(1+\lambda)}
{\lambda-\mu^*(1+0.62\lambda)}\right),
\label{McMillan}
\end{equation}
where $\mu^*$ is the Coulomb pseudopotential. Quite similar results were also
obtained with Allen-Dynes formula \cite{AD}, considered to be the best 
interpolation expression for $T_c$ in strongly coupled superconductors.
Taking the ``optimistic'' value of Coulomb pseudopotential $\mu^*=0$, we repeat 
our previous analysis. Now Eq.~(\ref{McMillan}) gives $\lambda$=0.61 for 
SrPt$_3$P and corresponding $T_c=5.6$K for LaPt$_3$P. Once we assume more
typical value of $\mu^*=0.1$, we get  $\lambda$=0.85  for SrPt$_3$P
and then $T_c$=5.4K for La compound is obtained. 

To sum up we see that the values of  $T_c$ for Sr and La systems reasonably
correlated with DOS behavior at Fermi level, which is similar to our 
estimates for iron pnictides and chalcogenides \cite{PvsC,ahight}. At the same
time, these estimates correspond to weak or intermediate coupling 
superconductivity in APt$_3$P superconductors, which can not explain 
(compare \cite{AD}) unusually
high $2\Delta/T_c$ ratios obtained from specific heat data in Ref. 
\cite{Takayama}. This stresses the need for independent experimental
estimates of $2\Delta/T_c$ ratios in these systems. 

\section {Conclusion}

In this work we presented LDA results for band dispersions, densities of states 
and Fermi surfaces for recently discovered Pt-pnictogen superconductors 
APt$_3$P (A=Sr,La) \cite{Takayama}. We confirm experimental predictions 
concerning the multiple band superconductivity in these systems, with 
complicated multiple sheet FS topology. 
In contrast to typical iron pnictides and chalcogenides we find Pt systems to be 
essentially three dimensional. Our LDA data and simple estimates of 
superconducting $T_c$ leave little space for strong coupling superconductivity
in new Pt-compounds. The observed correlation of $T_c$ values with DOS
behavior close to Fermi level stresses the importance of doping. It seems
probable, that hole doping of SrPt$_3$P, if possible, can lead to higher values
of $T_c$.

This work is partly supported by RFBR grant 11-02-00147 and was performed
within the framework of the Program of fundamental research of the Russian 
Academy of Sciences (RAS) ``Quantum mesoscopic and disordered structures'' 
(12-$\Pi$-2-1002).

\end{document}